\documentclass[english,12pt]{article}
\usepackage[cp1251]{inputenc}
\usepackage{babel}
\usepackage{amsfonts, amsmath}

\newcommand{\be}{\begin{equation}} \newcommand{\ee}{\end{equation}}

\newcommand{\gsim}{\mathrel{\hbox{\rlap{\lower.55ex\hbox{$\sim$}} \kern-.3em \raise.4ex \hbox{$>$}}}}

\begin{document}
\begin{center}
{\bf Minimal Length, Measurability and Gravity}\\
\vspace{5mm} A.E.Shalyt-Margolin \footnote{E-mail:
a.shalyt@mail.ru; alexm@hep.by; \linebreak Tel.:
+375-17-292-60-34}\\ \vspace{5mm} \textit{Research Institute for
Nuclear Problems, 11 Bobruiskaya str., Minsk 220040, Belarus}
\end{center}
 \rm\normalsize \vspace{0.5cm}
\begin{abstract}
The present work is a continuation of the previous papers written
by the author on the subject. In terms of the  measurability (or
measurable quantities) notion introduced in a minimal length
theory, first the consideration is given to a quantum theory  in
the momentum representation. The same terms are used to consider
the Markov gravity model that here illustrates the general
approach to studies of gravity in terms of  measurable quantities.
\\This  paper  is  dedicated to the 75th Anniversary of
Professor Vladimir  Grigor'evich Baryshevsky.
\end{abstract}

\section{Introduction: Measurable and Nonmeasurable Quantities}\label{Section 1}

This work is a direct continuation of my previous  papers
\cite{Shalyt-AHEP2,Shalyt-JAP1} and is interlaced with these
publications at some points.  As shown in \cite{Shalyt-AHEP2},
provided the theory involves the minimal length $l_{min}$ as
{\em{a minimal measurement unit}} for the quantities having the
dimensions of length, this theory must not have infinitesimal
spatial-temporal quantities $dx_{\mu}$ because the latter lead to
the infinitely small length $ds$ \cite{Einst1}
\begin{equation}\label{Introd 1}
ds^{2}=g_{\mu\nu}dx_{\mu}dx_{\nu}
\end{equation}
that is inexistent because of $l_{min}$.

Of course, in this case only   {\em{measurable}} quantities are
meant.  As a mathematical notion, the quantity $ds$ is naturally
existent but, due to the involvement of $l_{min}$, it is {
immeasurable}.

However it is well known that at high energies
(on the order of the quantum gravity energies) the minimal length
$l_{min}$ to which the indicated energies are ``sensitive'', as
distinct from the low ones, should inevitably become apparent in
the theory. But if $l_{min}$  is really present, it must be
present at all the ``Energy Levels'' of the theory, low energies
including.
 And this, in addition to the above arguments, points to the fact
that the mathematical formalism of the theory should not involve
any infinitesimal spatial-temporal quantities. Besides, some new
parameters become involved, which are dependent on $l_{min}$
\cite{QG1,Hawk-Penr1,Gar1,Planck2.1,Planck2.2,Planck2,Planck3.1,Planck3.2,Planck3,Planck4}.

What are the parameters of interest in the case under study?
It is obvious that, as the quantum-gravitational effects will be
revealed at very small (possibly Planck) scales, these parameters
should be dependent on some limiting values, e.g.,
 $l_P\propto  l_{min}$  and hence Planck energy $E_P$.

{\em{This means that in a high-energy gravitation theory the
energy-dependent or, what is the same, measuring scales-dependent
parameters should be necessarily introduced.}}

But, on the other hand, these parameters could hardly disappear totally at low energies,
\emph{i.e.}, for General Relativity (GR) too. However, since the
well-known canonical (and in essence the classical) statement of
GR has no such parameters \cite{Einst1}, the inference is as
follows: their influence at low energies is so small that it may
be disregarded at the modern stage in evolution of the theory and
of the~experiment.

{Still this does not imply that they should be ignored
in future evolution of the theory, especially on going to its
high-energy limit.}

But at the present time, the mathematical apparatus of both special and general
relativity theories (and of a quantum theory as well) is based on
the concept of continuity and on analysis of infinitesimal
spatial-temporal quantities. This is a corner stone for the
Minkowski space geometry (MS) and also for the Riemannian geometry
(RG) \cite{Einst1}.

However, this approach involves a problem when we proceed to a quantum description of nature.
 Even at a level of the heuristic understanding, it is clear that, as measuring procedures
 in a quantum theory are fundamental, the description with the use of infinitesimal
 quantities is problematic because in its character the measuring procedure is discrete.

 At a level of the mathematical formalism and physical principles, incompatibility
of both the Minkowski space geometry and Riemannian geometry with
the uncertainty principle is expected in any ``format'', in
relativistic and nonrelativistic cases. This problem is considered
in greater detail in the following sections of this work.

 Thus, if the matter concerns the {measurable} quantities only,
the Quantum Theory (QT) and Gravity formalism should be changed:
at least, a new formalism should not involve the infinitesimal
spatial-temporal quantities $dx_{\mu}$. Naturally, because of the
involved $l_{min}$ (initially assuming that $l_{min}\propto~l_P$)
new theories should involve new parameters associated with
$l_{min}$. Presently, such parameters are inexplicitly involved
(for example, $E/E_P$ in  a quantum  gravity phenomenology
\cite{QG1}).

 But there is no need to   discard the modern formalism of QT  and  Gravity,
since it is clear that at low energies it offers an excellent
approximation, experimentally supported to a high accuracy (see
\cite{Hawk-Penr1}). However, proceeding from the above, a
change-over to high energies is impossible as, by author's
opinion, this formalism is used in an effort to {combine
uncombinable things}.

 This work makes the arguments of \cite{Shalyt-AHEP2,Shalyt-JAP1} more forcible
on the one hand, and presents a study of the additional parameters
associated with the involvement of $l_{min}$ , in terms of which
one can develop a new formalism for a quantum theory and for
gravity at all the scales energies too, on the other hand.

One of the key problems of the modern fundamental physics
 (Quantum Theory (QT) and Gravity (GR)) is framing of a correct
  theory associated with the ultraviolet region,
\emph{i.e.}, the region of the highest (apparently Planck) energies
approaching those of the  Big Bang.

However, it is well known that at high energies (on the
order of the quantum gravity energies) the minimal length
$l_{min}$ to which the indicated energies are ``sensitive'', as
distinct from the low ones, should inevitably become apparent in
the theory. But if $l_{min}$  is really present, it must be
present at all the ``Energy Levels'' of the theory, low energies
including.

What follows from the existence of the minimal length
$l_{min}$? When the minimal length is involved, any nonzero {measurable}
quantity having the dimensions of length should be a multiple of $l_{min}$.
Otherwise, its {measurement} with the use of $l_{min}$ would result in the
{measurable} quantity $\varsigma$, so that $\varsigma<l_{min}$, and this is impossible.

 This suggests that the spatial-temporal quantities $dx_{\mu}$ are
{nonmeasurable} quantities because the latter lead to  the
infinitely small {nonmeasurable} quantity length $ds$
Equation~\eqref{Introd 1}.

Of course, as a mathematical notion, the quantities $dx_{\mu},ds$
are naturally existent but one should realize that there is no way
to express them in terms of the minimal possible measuring unit
$l_{min}$.

So, trying to frame a theory (QT and GR) correct at all the energy
levels using only the {measurable} quantities, one should realize
that then the mathematical formalism of the theory should not
involve any infinitesimal spatial-temporal quantities. Besides,
proceeding from the acknowledged results associated with the
Planck scales physics
\cite{QG1,Hawk-Penr1,Gar1,Planck2.1,Planck2.2,Planck2,Planck3.1,Planck3.2,Planck3,Planck4},
one can infer that certain new parameters dependent on $l_{min}$
should be involved.

 What are the parameters of interest in the case under study?
It is obvious that, as the quantum-gravitational effects will be
revealed at very small (possibly Planck) scales, these parameters
should be dependent on some limiting values, e.g.,
 $l_P\propto  l_{min}$  and hence Planck's energy~$E_P$.

{This means that in high-energy QT and GR the energy-or,
what is the same, measuring scales-dependent parameters should be
necessarily introduced.}

But, on the other hand, these parameters could hardly disappear totally
at low energies both in QT and in GR.

But, provided $l_{min}$  exists, it must be involved at all the energy levels,
both  {high} and {low}.

The fact that $l_{min}$ is omitted in the formulation of low-energy
QT and GR and the theories give perfect results leads to two different inferences:
\\
\\{\bf 1.} The influence of the above-mentioned new parameters
associated with  $l_{min}$ in low-energy QT  and  GR is so small
that it may be disregarded at the modern stage in evolution of the
theory and of the experiment.
\\
\\{\bf 2.} The modern mathematical apparatus of conventional QT and GR has been
derived in terms of the   infinitesimal spatial-temporal
quantities $dx_{\mu}$ which, as noted above,  are { nonmeasurable
quantities} in the formalism of $l_{min}$.

\section{Main Motivation}

In this Section the principal assumptions are introduced which
have been implicitly used previously in \cite{Shalyt-AHEP2} and
especially in \cite{Shalyt-JAP1}.

It is well known that in a quantum study the key role is played by
the measuring procedure, its fundamental principle being  the
Heisenberg Uncertainty Principle (HUP) \cite{Heis1,Mess}:
\begin{equation}\label{HUP1}
\Delta x\geq\frac{\hbar}{\Delta p}
\end{equation}

(Note that the normalization $\Delta x \Delta p\geq \hbar$ is used
rather than $\triangle x \Delta p\geq \hbar/2$.)

 Now we can proceed to the following quite natural suppositions.\vspace{6pt}

\noindent \textbf{Supposition 1.} {Any small variation (increment) $\widetilde{\Delta} x_{\mu}$
of any spatial coordinate $x_{\mu}$ of the arbitrary point
$x_{\mu},\mu~=~1,...,3$ in some space-time system $\emph{R}$ may
be realized in the form of the uncertainty (standard deviation)
$\Delta x_{\mu}$ when this coordinate is measured within the scope
of Heisenberg's Uncertainty Principle (HUP)
\begin{equation}\label{HUP01}
\widetilde{\Delta} x_{\mu}=\Delta x_{\mu},\Delta x_{\mu}\simeq
\frac{\hbar}{\Delta p_{\mu}}, \mu=1,2,3
\end{equation}
for some $\Delta p_{\mu}\neq 0$.

Similarly, for $\mu=0$ for pair ``time-energy'' $(t,E)$, the any small variation (increment)
 value of time $\widetilde{\Delta} x_{0}=\widetilde{\Delta} t_{0}$ may
be realized in the form of the uncertainty (standard deviation)
$\Delta x_{0}=\Delta t$ and~then
\begin{equation}\label{HUP02}
\widetilde{\Delta} t=\Delta
t,\Delta t\simeq \frac{\hbar}{\Delta E}
\end{equation}
for some $\Delta E\neq 0$}.

Here HUP is given for the nonrelativistic case. In the relativistic case HUP
has the distinctive features \cite{Land2} which, however, are of
no significance for the general formulation of {Supposition~1},
being associated only with particular alterations in the
right-hand side of the second relation Equation~\eqref{HUP01} as shown
later.

It is clear that at low energies $E\ll E_P$ (momentums $P\ll P_{pl}$) {Supposition 1}
sets a lower bound for the variations (increments)
$\widetilde{\Delta} x_{\mu}$  of any space-time coordinate
$x_{\mu}$.

At high energies $E$ (momentums $P$) this is not the case if
$E$ ($P$) have no upper limit. But, according to the modern
knowledge, $E$ ($P$) are bounded by some maximal quantities
$E_{max}$, ($P_{max}$)
\begin{equation}\label{HUP03}
E\leq E_{max},P\leq P_{max},
\end{equation}
where in general   $E_{max},P_{max}$  may be on the order of
Planck quantities $E_{max}\propto E_P,P_{max}\propto P_{pl}$ and
also may be the trans-Planck's quantities.

In any case the quantities $P_{max}$  and $E_{max}$ lead
to the introduction of the minimal length $l_{min}$ and of the
minimal time $t_{min}$.

With this point of view, even at the ultimate (Planck) energies a minimal
``detected'' (\emph{i.e.}, measurable) space-time volume is, within the
known constants, restricted to
\begin{equation}\label{GUP1.new3.2}
V_{min}\propto l^{4}_{P}.
\end{equation}

Consequently, ``detectability'' of the infinitesimal space-time
volume
\begin{equation}\label{GUP1.new3.3}
V_{dx_{\mu}}= (dx_{\mu})^{4}
\end{equation}
is impossible as this necessitates going to infinitely high
energies
\begin{equation}\label{GUP1.new3.4}
E\rightarrow \infty.
\end{equation}

Because of this, it is natural to complete Supposition 1 with
Supposition 2.\vspace{6pt}

\noindent \textbf{Supposition 2.} {There is the minimal length $l_{min}$  as {a minimal
measurement unit} for all quantities having the dimension of
length, whereas the minimal time  $t_{min}=l_{min}/c$  as {a
minimal measurement unit} for all quantities having the dimension
of time, where $c$ is the speed of light.}

$l_{min}$ and $t_{min}$ are naturally introduced as $\Delta
x_{\mu},\mu=1,2,3$  and $\Delta t$  in Equations~\eqref{HUP01} and (\ref{HUP02})
for $\Delta p_{\mu}=P_{max}$ and $\Delta E=E_{max}$.

For definiteness, we consider that $E_{max}$ and $P_{max}$
are the quantities on the order of the Planck quantities, then
$l_{min}$  and $t_{min}$ are also on the order of Planck
quantities $l_{min}\propto l_P$, $t_{min}\propto t_P$.

{Suppositions 1 and 2} are quite natural in the sense that there
 are no physical principles with which these suppositions are inconsistent.

\section{Minimal Length and Measurability}
\label{Section3}

In this Section particularly the results from Subsection 3.1 of
\cite{Shalyt-JAP1} are used. Now from the start we assume that the
theory involves the minimal length $l_{min}$ as {a minimal
measurement unit} for all quantities having the dimension of
length.

Then it is convenient to begin our study not with HUP Equation~\eqref{HUP1}
but with its widely known high-energy generalization---the {
Generalized Uncertainty Principle (GUP)} that naturally leads to
the minimal length $l_{min}$ \cite{Ven1,Ven2,Ven3,Polch,GUPg1,Ahl1,Ahl,Magg1,Magg2,Magg3,Capoz,Kempf,Nozari}:
\begin{equation}\label{GUP1}
\Delta x\geq\frac{\hbar}{\Delta p}+\alpha^{\prime}
l_{P}^2\frac{\Delta p}{\hbar}.
\end{equation}

Here $\alpha^{\prime}$ is the model-dependent dimensionless
numerical factor   and  $l_{P}$  is the Planckian length.
\\Note also that initially GUP Equation~\eqref{GUP1} was derived within a string theory
\cite{Ven1,Ven2,Ven3,Polch} and, subsequently, in a series of works
far from this theory \cite{GUPg1,Ahl1,Ahl,Magg1,Magg2,Magg3,Capoz} it has been
demonstrated that on going to high (Planck) energies in the
right-hand side of HUP Equation~\eqref{HUP1} an additional ``high-energy''
term  $\propto l_{P}^2\frac{\triangle p}{\hbar}$ appears. Of
particular interest is the work \cite{GUPg1}, where by means of a
simple gedanken experiment it has been demonstrated that with
regard to the gravitational interaction Equation~\eqref{GUP1}  is the case.

As Equation~\eqref{GUP1} is a quadratic inequality, then it naturally leads to
the minimal length \mbox{$l_{min}= \xi l_P=2\surd \alpha^{\prime}l_P$.}

This means that the theory for the quantities with a particular dimension
has a {minimal measurement unit}. At least, all the quantities
with such a dimension should be ``quantized'', \emph{i.e.}, be measured by
an integer number of these {minimal units} of measurement.

 Specifically, if   $l_{min}$---{minimal unit} of length, then for
any length  $L$ we have the {``Integrality Condition'' (IC)}
 \begin{equation}\label{Introd 2.4}
 L=N_{L}l_{min},
 \end{equation}
where~$N_{L}~>~0$~is an integer number.

What are the consequences for GUP Equation~\eqref{GUP1} and HUP Equation~\eqref{HUP1}?

Assuming that HUP is to a high accuracy derived from GUP on going to low energies
or that  HUP is a special case of  GUP  at low values of the
momentum, we have
\begin{equation}\label{GUP2} (GUP,\Delta p
\rightarrow 0)=(HUP).
\end{equation}

By the language of $N_{L}$  from Equations~\eqref{Introd 2.4} and (\ref{GUP2})
is nothing else but a change-over to the~following:
 \begin{equation}\label{GUP3}
 (N_{\Delta x}\approx 1)\rightarrow (N_{\Delta x}\gg 1).
\end{equation}

The assumed equalities in Equations~\eqref{HUP1} and (\ref{GUP1}) may be
conveniently rewritten in terms of $l_{min}$ with the use of the
deformation parameter $\alpha_{a}$. This parameter has been
introduced earlier in the papers  \cite{shalyt1,shalyt2,shalyt3,shalyt4,shalyt5,shalyt6,shalyt7,shalyt9}
as a {deformation parameter} on going from the canonical
quantum mechanics to  the quantum mechanics at Planck's scales
(early Universe) that is considered to be the quantum mechanics
with the minimal length (QMML):
\begin{equation}\label{D1}
\alpha_{a}=l_{min}^{2}/a^{2},
\end{equation}
where $a$ is the measuring scale.
\\
\\{\bf Definition 1}
\\{\it Deformation is understood as an
extension of a particular theory by inclusion of one or several
additional parameters in such a way that the initial theory
appears in the limiting transition} \cite{Fadd}.
\\
\\Then with the equality ($\Delta p\Delta x= \hbar$)
Equation~\eqref{GUP1} is of the form
\begin{equation}\label{GUP1.new1}
\Delta x=\frac{\hbar}{\Delta p}+\frac{\alpha_{\Delta x}}{4}\Delta
x.
\end{equation}

In this case due to Equations (\ref{Introd 2.4}), (\ref{GUP3}) and
(\ref{GUP1.new1}) takes the following form:
\begin{equation}\label{GUP1.new2}
N_{\Delta x}l_{min}=\frac{\hbar}{\Delta p}+\frac{1}{4N_\Delta x
}l_{min}
\end{equation}
or
\begin{equation}\label{GUP1.new3}
(N_{\Delta x}-\frac{1}{4N_\Delta x })l_{min}=\frac{\hbar}{\Delta
p}.
\end{equation}

That is
\begin{equation}\label{GUP1.new4.1}
\Delta p=\frac{\hbar}{(N_{\Delta x}-\frac{1}{4N_\Delta
x})l_{min}}.
\end{equation}

From Equations~\eqref{GUP1.new2}--(\ref{GUP1.new4.1}) it is clear
that HUP Equation (\ref{HUP1}) in the case of the equality appears
to a high accuracy in the limit $N_\Delta x\gg 1$ in conformity
with Equation~\eqref{GUP3}.

 It is easily seen that the parameter $\alpha_{a}$  from Equation~\eqref{D1}
is discrete as it is nothing else but
\begin{equation}\label{D1.new1}
\alpha_{a}=l_{min}^{2}/a^{2}=\frac{l_{min}^{2}}{N^{2}_{a}l_{min}^{2}}=\frac{1}{N^{2}_{a}}.
\end{equation}

At the same time, from Equation~\eqref{D1.new1} it is evident that
$\alpha_{a}$ is irregularly discrete.

It is clear that from Equation (\ref{GUP1.new4.1}) at low energies
($N_\Delta x\gg 1$), up to a constant
\begin{equation}\label{GUP1.new5.1}
\frac{\hbar^{2}}{l_{min}^{2}}=\frac{\hbar
c^{3}}{4\alpha^{\prime}G}
\end{equation}
we have
\begin{equation}\label{GUP1.new5}
\alpha_{\Delta x}=(\Delta p)^{2}.
\end{equation}

But all the above computations are associated with the
nonrelativistic case. As known, in the relativistic case, when the
total energy of a particle with the mass   $m$ and with the
momentum $p$ equals \cite{Land1}:
\begin{equation}\label{GUP1.new6}
E=\sqrt {{p^{2}}c^{2}+m^{2}c^{4}},
\end{equation}
a minimal value for $\Delta x$  takes the form \cite{Land2}:
\begin{equation}\label{GUP1.new7}
\Delta x\approx \frac{c \hbar}{E}.
\end{equation}

And in the {ultrarelativistic case}
\begin{equation}\label{GUP1.new7.1}
E\approx pc
\end{equation}
this means simply that
\begin{equation}\label{GUP1.new7.2}
\Delta x\approx \frac{\hbar}{p}.
\end{equation}

Provided the minimal length $l_{min}$ is involved and considering
the {``Integrality Condition'' (IC)}  Equation~\eqref{Introd 2.4}, in the
general case for Equation~\eqref{GUP1.new7} at the energies considerably
lower than the Planck energies $E\ll E_{P}$ we obtain the
following:
\begin{eqnarray}\label{GUP1.new8}
\Delta x=N_{\Delta x}l_{min}\approx \frac{c \hbar}{E},\nonumber
\\or
\nonumber
 \\E\approx \frac{c \hbar}{N_{\Delta x}}.
\end{eqnarray}

Similarly, at the same energy scale in the ultrarelativistic case
we have
\begin{eqnarray}\label{GUP1.new8.1}
p\approx \hbar/ N_{\Delta x}.
\end{eqnarray}

Next under {Supposition 2}, we assume that  there is a
minimal measuring unit of time
\begin{equation}\label{Min.time}
t_{min}=l_{min}/v_{max} = l_{min}/c.
\end{equation}

Then the foregoing Equations (\ref{HUP1})--(\ref{GUP1.new3}) are
rewritten by substitution as follows:
\begin{equation}\label{GUP1.new4}
\Delta x\rightarrow \Delta t; \Delta p\rightarrow \Delta E;
l_{min}\rightarrow t_{min};N_{L}\rightarrow N_{t=L/c}
\end{equation}

Specifically, Equation~\eqref{GUP1.new3} takes the form
\begin{equation}\label{GUP1.new3.1}
(N_{\Delta t}-\frac{1}{4N_\Delta t })t_{min}=\frac{\hbar}{\Delta
E}.
\end{equation}

And similar to Equation~\eqref{Introd 2.4}, we get the {``Integrality
Condition'' (IC)} for any time $t$:
 \begin{equation}\label{Introd 2.4new}
t\equiv t(N_{t})=N_{t}t_{min},
\end{equation}
for certain an integer $|N_{t}|~\geq~0$.

According to Equation~\eqref{GUP1.new3.1}, let us define the corresponding energy $E$
 \begin{equation}\label{Introd 2.4new2}
E\equiv E(N_{t})=\frac{\hbar}{|N_{t}-\frac{1}{4N_t}|t_{min}}.
 \end{equation}

Note that at low energies $E\ll E_P$, that is for
$|N_{t}|\gg 1$, the formula Equation~\eqref{Introd 2.4new2} naturally takes the following form:
 \begin{equation}\label{Introd 2.4new3}
E\equiv E(N_{t})=\frac{\hbar}{|N_{
t}|t_{min}}=\frac{\hbar}{|t(N_{t})|}.
 \end{equation}

{\bf Definition 2} ({\textbf{Measurability}})
\\(1) {\it Let us define
the quantity having the dimensions of length $L$ or time $t$
{measurable}, when it satisfies the relation
Equation~\eqref{Introd 2.4} (and respectively
Equation~\eqref{Introd 2.4new}).}
\\(2){\it Let us define any physical quantity {measurable},
 when its value is consistent with points (1)  of this
 Definition.}
\\
\\Thus, {infinitesimal changes} in length (and hence in time) are
{impossible} (to that indicated in Section \ref{Section 1}) and
any such changes are dependent on the existing energies.

In particular, a minimal possible  {measurable}  change of length
is $l_{min}$. It corresponds to some maximal value of the energy
$E_{max}$ or momentum $P_{max}$, If $l_{min}\propto l_P$, then
$E_{max}\propto E_P$,$P_{max}\propto P_{Pl}$, where
$P_{max}\propto P_{Pl}$, where $P_{Pl}$ is where the Planck
momentum.  Then denoting in {nonrelativistic} case with
$\triangle_{p}(w)$ a {minimal measurable} change every spatial
coordinate $w$ corresponding to the energy $E$ we~obtain
 \begin{equation}\label{Min1.1}
\triangle_{P_{max}}(w)=\triangle_{E_{max}}(w)=l_{min}.
\end{equation}

Evidently, for lower energies (momentums) the corresponding values
of $\triangle_{p}(w)$ are higher and, as the quantities having the
dimensions of length are quantized
 Equation~\eqref{Introd 2.4}, for $p\equiv p(N_{p})<p_{max}$, $\triangle_{p}(w)$ is transformed to
 \begin{equation}\label{Min2}
 |\triangle_{p(N_{p})}(w)|=|N_{p}|l_{min}.
 \end{equation}
where~$|N_{p}|~>~1$~is an integer  number  so that we have
\begin{equation}\label{Min3}
|N_{p}-\frac{1}{4N_{p}}|l_{min}=\frac{\hbar}{|p(N_{p})|}.
 \end{equation}

In the relativistic case the Equation (\ref{Min1.1}) holds,
whereas
 Equations~\eqref{Min2} and (\ref{Min3}) for $E\equiv
 E(N_{E})<E_{max}$ are replaced by
\begin{equation}\label{Min2.1}
 |\triangle_{E(N_{E})}(w)|=|N_{E}|l_{min},
 \end{equation}
where $|N_{E}| > 1$ is an integer.

Next we assume that at high energies $E\propto E_P$ there is a
possibility only for the {nonrelativistic} case or {
ultrarelativistic} case.

 Then for the {ultrarelativistic} case, with regard
to Equations~\eqref{GUP1.new7.1}--(\ref{GUP1.new3.1}), Formula (\ref{Min3})
takes the~form
\begin{equation}\label{Min3.1}
|N_{E}-\frac{1}{4N_{E}}|l_{min}=\frac{\hbar
c}{E(N_{E})}=\frac{\hbar}{|p(N_{p})|},
 \end{equation}
where $N_{E}=N_{p}$.

 In the relativistic case at low energies we have
\begin{equation}\label{Min2.1}
 E\ll E_{max}\propto E_P.
\end{equation}

 In accordance with Equations~\eqref{GUP1.new6} and (\ref{GUP1.new7}) and
Formula (\ref{Min2})  is of the form
\begin{equation}\label{Min3.new}
|\triangle_{E(N_{E})}(w)|=|N_{E}|l_{min}=\frac{\hbar
c}{E(N_{E})},|N_{E}|\gg~1~-~integer.
 \end{equation}

In the nonrelativistic case at low energies Equation~\eqref{Min2.1} due to
Equation~\eqref{Min3} we get
\begin{equation}\label{Min4}
|\triangle_{p(N_{p})}(w)|=|N_{p}|l_{min}=\frac{\hbar}{|p(N_{p})|},|N_{p}|\gg
1-integer.
 \end{equation}

 In a similar way for the time coordinate $t$, by virtue of Equations (\ref{Introd
2.4new})--(\ref{Introd 2.4new3}), at the same conditions
 we have  similar Equations (\ref{Min1.1})--(\ref{Min3})
\begin{eqnarray}\label{Min5}
\triangle_{E_{max}}(t)=t_{min}.
\end{eqnarray}

For $E\equiv E(N_{t})<E_{max}$
\begin{equation}\label{Min2.new1}
 |\triangle_{E(N_{t})}(t)|=|N_{t}|t_{min},
 \end{equation}
 where $|N_{E}|>1$ is an integer,
so that we obtain
\begin{equation}\label{Min3.new1}
|N_{t}-\frac{1}{4N_{t}}|t_{min}=\frac{\hbar c}{E(N_{t})}.
 \end{equation}

 In the relativistic case at low energies
\begin{equation}\label{Min2.1new}
 E\ll E_{max}\propto E_P,
\end{equation}
in accordance with Equations~\eqref{GUP1.new6} and
(\ref{GUP1.new7}), Equation (\ref{Min2})  takes the form
\begin{equation}\label{Min3.new2}
|\triangle_{E(N_{t})}(w)|=|N_{t}|l_{min}=\frac{\hbar
c}{E(N_{t})},|N_{t}|\gg 1-integer.
\end{equation}

{\bf Remark 1.}
\\
\\{\bf 1.1.} It should be noted that the lattice is
usually understood as a uniform discrete structure with one and
the same constant parameter  $a$  (lattice pitch). But in this
case we have a nonuniform discrete structure (lattice in its
nature), where the analogous parameter is variable, is a multiple
of  $l_{min}$, \emph{i.e.}, $a=N_{a}l_{min}$, and also is
dependent on the energies.  Only in the limit of high (Planck's)
energies we get a (nearly) uniform lattice with (nearly) constant
pitch $a\approx l_{min}$  or $a=\kappa l_{min}$ where $\kappa$ is
on the order of 1.
\\
\\{\bf 1.2.} Obviously, when $l_{min}$ is involved, the
foregoing formulas for the momenta $p(N_{p})$  and for the
energies $E(N_{E}),E(N_{t})$ may {certainly} give the highly
accurate result that  is close to the experimental one only at the
verified low energies:  $|N_{p}|\gg 1,|N_{E}|\gg 1,|N_{t}|\gg 1$.
In the case of high energies   $E\propto E_{max}\propto E_P$ or,
what is the same $|N_{p}|\rightarrow 1,|N_{E}|\rightarrow
1,|N_{t}|\rightarrow 1$, we have a certain, experimentally
unverified, model with a correct low-energy limit.
\\
\\{\bf 1.3.} It should be noted that dispersion relations
Equation~\eqref{GUP1.new6} are valid only at low energies  $E\ll
E_P$. In the last few years in a series of works
\cite{Hoss,Tawf,Vagen,Nozar1,Nozar2} it has been demonstrated that
within the scope of GUP the high-energy generalization of
Equation~\eqref{GUP1.new6}---Modified Dispersion Relations
(MDRs)---is valid.

Specifically, in its most general form the Modified Dispersion Relation (Formula (9) in \cite{Nozar2})
may be given as follows:
\begin{equation}
\label{math:1.5}
 p^2 = f(E,m;l_p) \simeq E^2 - \mu^2+ \alpha_1 l_p E^3
+ \alpha_2 l_p^2 E^4+ O\left(l_p^3 E^5 \right),
\end{equation}
where in the notation of \cite{Nozar2} the fundamental constants
are $c=\hbar=k_{B}=1$, $f$ is the function that gives the exact
dispersion relation, and in the right-hand side the applicability
of the Taylor-series expansion for $E \ll 1/l_P$ is assumed. The
coefficients $\alpha_i$ can take different values in different
quantum-gravity proposals.  $m$ is the rest energy of a particle,
and the mass parameter $\mu$ in the right-hand side is directly
related to the rest energy but $\mu \neq m$ if not all the
coefficients $\alpha_i$ are~vanishing.

The general case of (MDRs) Equation~\eqref{math:1.5} in terms of the considerations given
in this section is yet beyond the scope of this paper and
necessitates further studies of the transition from low $E\ll E_P$
to high $E\approx E_P$ energies.

 For now it is assumed that at low energies Equation (\ref{GUP1.new6})
 is valid to within a high accuracy, whereas at high energies, \emph{i.e.}, for
$|N_{p}|\rightarrow 1,|N_{E}|\rightarrow 1,|N_{t}|\rightarrow 1$,
Equation~\eqref{GUP1.new6}  should be replaced by
Equation~\eqref{math:1.5}. Besides, it is important that in this
paper, as distinct from \cite{Hoss,Tawf,Vagen,Nozar1,Nozar2}, the
author uses the simplest (earlier) variant of GUP
\cite{Ven1,Ven2,Ven3,Polch,GUPg1,Ahl1,Ahl,Magg1,Magg2,Magg3,Capoz},
involving a minimal length but not a minimal momentum.

Also note that references   \cite{Hoss,Tawf,Vagen,Nozar1,Nozar2}
give not nearly so complete a list of the publications devoted to
GUP (and, in particular, MDR)---a very complete and interesting
survey may be found in \cite{Tawf}.
\\
\\{\bf 1.4.} The papers \cite{Shalyt-AHEP2,Shalyt-JAP1}
point to the fact that the resolved discrete theory  is very close
to the initial continuous one ($l_{min}=0$) at low energies $E\ll
E_P$, \emph{i.e.}, at $|N_{p}|\gg 1,|N_{E}|\gg 1$.
\\
\\In what follows all the considerations are given in terms of
{``measurable quantities''} in the sense of {Definition 2}  given
in this Section. Specifically, in Section \ref{Section5} these
terms are used to consider the {Momentum Representation} for a
quantum theory.

\section{Space-Time Lattice of Measurable Quantities and Dual
Lattice}
\label{Section4}

So, provided the minimal length $l_{min}$  exists, two lattices
are naturally arising.
\\{\bf I.} Lattice of the {space-time variation}---$Lat_{S-T}$ representing,
 to within the known multiplicative constants, the sets  of nonzero
integers $N_{w}\neq 0$ and $N_{t}\neq 0$ in the corresponding
formulas from the set Equations~\eqref{Min2} and (\ref{Min3.new2}) for each of
the three space variables $w\doteq x;y;z$ and the time  variable~$t$

\begin{equation}\label{Latt1}
Lat_{S-T}\doteq (N_{w},N_{t}).
\end{equation}

Which restrictions should be initially imposed on these sets of
nonzero integers?

 It is clear that in every such set all the integers $(N_{w},N_{t})$
should be sufficiently ``close'', because otherwise, for one and the
same  space-time point, variations in the values of its different
coordinates are associated with principally different values of
the energy  $E$ which are ``far'' from each other.

 Note that the words ``close'' and ``far'' will be elucidated further in this text.

 Thus, at the admittedly low energies (Low Energies) $E\ll E_{max}\propto
E_P$ the low-energy part (sublattice) $Lat_{S-T}[LE]$ of
$Lat_{S-T}$ is as follows:
\begin{equation}\label{Latt2}
Lat_{S-T}[LE]= (N_{w},N_{t})\equiv (|N_{x}|\gg 1,|N_{y}|\gg
,|N_{z}|\gg 1,|N_{t}|\gg 1).
\end{equation}
At high energies (High Energies) $E\rightarrow E_{max}\propto E_P$
we, on the contrary, have the sublattice $Lat_{S-T}[HE]$ of
$Lat_{S-T}$
\begin{equation}\label{Latt3}
Lat_{S-T}[HE]= (N_{w},N_{t})\equiv (|N_{x}|\approx
1,|N_{y}|\approx1,|N_{z}|\approx 1,|N_{t}|\approx 1).
\end{equation}

{\bf II.} Next let us define the lattice {momenta-energies
variation} $Lat_{P-E}$
 as a set to {obtain} $(p_{x}(N_{x,p}),p_{y}(N_{y,p}),p_{z}(N_{z,p}),E(N_{t}))$
in the nonrelativistic and ultrarelativistic cases for all
energies, and as a set to {obtain}
\\$(E_{x}(N_{x,E}),E_{y}(N_{y,E}),E_{z}(N_{z,E}),E(N_{t}))$ in the
relativistic
\\(but not ultrarelativistic) case for low energies
$E\ll E_P$, where all the components of the above sets conform to
the space coordinates $(x,y,z)$  and time coordinate $t$ and are
given by the corresponding Formulas
(\ref{Min1.1})--(\ref{Min3.new2}) from the previous Section.

 Note that, because of the suggestion made after formula
Equation~\eqref{Min2.1} in the previous Section, we can state that the
foregoing sets exhaust all the collections of momentums and
energies possible for the lattice $Lat_{S-T}$.
\\ From this it is inferred that, in analogy with point
I of this Section, within the known multiplicative constants, we
have
\begin{equation}\label{Latt1.1}
Lat_{P-E}\doteq
(\frac{1}{N_{w}-\frac{1}{1/4N_{w}}},\frac{1}{N_{t}-\frac{1}{1/4N_{t}}}),
\end{equation}
where $N_{w}\neq 0, N_{t}~\neq~0$~are~integer numbers from
Equation~\eqref{Latt1}. Similar to Equation~\eqref{Latt2}, we
obtain the low-energy (Low Energy) part or the sublattice
$Lat_{P-E}[LE]$ of $Lat_{P-E}$
\begin{equation}\label{Latt2.1}
Lat_{P-E}[LE]\approx(\frac{1}{N_{w}},\frac{1}{N_{t}}),|N_{w}|\gg
1,|N_{t}|\gg 1.
\end{equation}

In accordance with Equation~\eqref{Latt3}, the high-energy (High Energy)
part (sublattice) $Lat_{P-E}[HE]$ of $Lat_{P-E}$ takes the form
\begin{equation}\label{Latt3.1}
Lat_{P-E}[HE]\approx
(\frac{1}{N_{w}-\frac{1}{1/4N_{w}}},\frac{1}{N_{t}-\frac{1}{1/4N_{t}}}),
|N_{w}|\rightarrow 1,|N_{t}|\rightarrow 1.
\end{equation}

Considering {\bf Remark 1}  from the previous Section, it should
be noted that, as currently the low energies $E\ll E_{max}\propto
E_P$  are verified by numerous experiments and thoroughly studied,
the sublattice $Lat_{P-E}[LE]$ Equation~\eqref{Latt2.1} is
correctly defined and rigorously determined by the sublattice
$Lat_{S-T}[LE]$ Equation~\eqref{Latt2}.

However, at high energies $E\rightarrow E_{max}\propto
E_P$ we can not be so confident the sublattice $Lat_{P-E}[HE]$ may
be defined more exactly.

Specifically, $\alpha_{a}$ is obviously a small parameter.
And, as demonstrated in
\cite{shalyt-entropy2,shalyt-IJMPD}, in the case of  GUP we
have the following:
\begin{equation} \label{comm5}
[\vec{x},\vec{p}]=i\hbar(1+a_{1}\alpha+a_{2}\alpha^{2}+...).
\end{equation}

But, according to Equation~\eqref{D1.new1}, $|1/N_{a}|=\sqrt{\alpha_{a}}$,
then, due to  Equation~\eqref{comm5}, the denominators in the right-hand
side of Equation~\eqref{Latt3.1} may be also varied by adding the terms
$\propto 1/N^{2}_{w},\propto 1/N^{3}_{w}...$,$\propto
1/N^{2}_{t},\propto 1/N^{3}_{t}...$, that is liable to influence
the final result for $|N_{w}|\rightarrow 1,|N_{t}|\rightarrow 1$.
\\ The notions ``close'' and ``far'' for $Lat_{P-E}$
will be completely determined by the dual lattice  $Lat_{S-T}[LE]$
and by Formulas (\ref{Min2}) and (\ref{Min3.new2}).

It is important to note the following.

{In the low-energy sublattice $Lat_{P-E}[LE]$
all elements are varying very smoothly enabling the approximation
of a continuous theory}.

\section{Measurable Quantities and Momentum Representation}
\label{Section5}

For convenience, we denote the minimal length $l_{min}\neq 0$ by $\ell$.

Let us consider the above calculations using
the formalism of the well-known work \cite{Kempf}. Then GUP
(Section
 3.2 in \cite{Kempf}) has the following form:
\begin{equation} \label{UR1}
[{\bf{x}},{\bf{p}}] = i\hbar (1 + \beta {\bf{p}}^2),
\end{equation}
where ($\beta >0$) and
\begin{equation}\label{UR2}
 \beta=\frac{\ell^{2}}{
\hbar^{2}}.
\end{equation}

In the form of Section \ref{Section3} in the present work, Formula (7) from
\cite{Kempf}
\begin{equation}\label{UR3}
\Delta x \Delta p \ge \hbar (1+\beta (\Delta p)^2 + \beta \langle
{\bf{p}} \rangle^2)
\end{equation}
with regard to Equations~\eqref{Introd
2.4}, (\ref{GUP1.new2}), (\ref{GUP1.new4.1}) and (\ref{UR2}) may be written as
\begin{equation}\label{UR3.1}
\frac{\hbar N_{\Delta x}}{(N_{\Delta x}-\frac{1}{4N_\Delta x})}\ge
\hbar (1+\frac{1}{(N_{\Delta x}-\frac{1}{4N_\Delta x})^{2}}   +
\frac{\ell^{2}}{ \hbar^{2}} \langle {\bf{p}} \rangle^2).
\end{equation}

In the equality case this results in the following expression:
\begin{equation}\label{UR3.2}
\frac{-\hbar^{2}(12N_{\Delta x}^{2}+1)}{(4N_{\Delta
x}^{2}-1)^{2}\ell^{2}}=
\frac{-\hbar^{2}}{\ell^{2}}(3+\frac{4}{(4N_{\Delta
x}^{2}-1)^{2}})= \langle {\bf{p}}\rangle^2.
\end{equation}

In this way at low energies $E\ll E_P$, \emph{i.e.}, at $|N_{\Delta
x}|\gg 1$, $\langle {\bf{p}}\rangle^2$  is varying practically continuously.
\\ Next, hereinafter we use the Formula (\ref{Min3}) with the replacement of $l_{min}=\ell$, \emph{i.e.}, we have
\linebreak $N_{\Delta x}=N_{p}$~and
\begin{equation}\label{UR3.3.1}
|p_{N}|=\frac{\hbar}{|N_{p}-\frac{1}{4N_{p}}|\ell}.
\end{equation}

We can write
\begin{eqnarray}\label{UR3.3-new2}
\imath \hbar(1+\beta
p^2)=\imath\hbar(1+\frac{\ell^{2}}{\hbar^{2}}\frac{\hbar^{2}}{(N_{p}-\frac{1}{4N_{p}})^{2}\ell^{2}})=\imath\hbar
(1+\frac{1}{(N_{p}-\frac{1}{4N_{p}})^{2}}).
\end{eqnarray}

Let us introduce the following symbols:
\begin{eqnarray}\label{UR3.3-new2.1}
\Delta_{p}p_{N}=p_{N}-p_{N+1};\Delta^{-1}_{p}\psi(p_{N})=\frac{\psi(p_{N})-\psi(p_{N+1})}{p_{N}-p_{N+1}}=
\nonumber
\\=\frac{\psi(p_{N+1}+\Delta_{p}p_{N})-\psi(p_{N+1})}{\Delta_{p}p_{N}}.
\end{eqnarray}

Then we suppose that only in the {classical dynamics}
{variations} of momenta (energies) have no lower bounds and we
have $dp$. At the same time, in a {quantum dynamics}, due to
the limited spatial domains, these {variations} have both
upper and lower bounds.

 In this case, as distinct from \cite{Kempf},
in the theory there is a {minimum variation of the momentum}
$\Delta p_{min}$ that within the scope of  the {measurability}
({Definition 2} in Section \ref{Section3}) takes the form
\begin{equation}\label{UR3.3}
\Delta p_{min}\equiv \textmd{p}=\frac{\hbar}{\ell}
\frac{1}{(\textbf{N}-\frac{1}{4\textbf{N}})}\approx\frac{\hbar}{\ell\textbf{N}}.
\end{equation}

As in Equation~\eqref{UR3.3-new2.1} at high $|N_{p}|,(|N_{p}|\gg 1)$,
$\Delta_{p}p_{N}=p_{N}-p_{N+1}\propto
(\frac{1}{N_{p}}-\frac{1}{N_{p}+1})=\frac{1}{N_{p}(N_{p}+1)}$, it is clear that
\begin{eqnarray}\label{UR3.3-new2.2}
N_{p}(N_{p}+1)\leq \textbf{N}\enskip or \enskip -\frac{1}{2}-\sqrt
(\frac{1}{4}+\textbf{N})\leq N_{p}\leq-\frac{1}{2}+\sqrt
(\frac{1}{4}+\textbf{N}).
\end{eqnarray}

Considering that $N_{p}$ is an integer number and $\textbf{N}\gg
1$, it follows that
\begin{eqnarray}\label{UR3.3-new2.3}
|N_{p}|\leq [\sqrt \textbf{N}]-1\equiv \widetilde{\textbf{N}},
\end{eqnarray}
where the square brackets $[\enskip]$ in the right-hand side  of
Equation~\eqref{UR3.3-new2.3} denote an integer part of the~number.

Next, due to Equations~\eqref{UR3.3-new2} and (\ref{UR3.3-new2.1}),
an analog of Formulae  (11) and (12) from \cite{Kempf} in the case
under study {at low energies} will be of the form
\begin{eqnarray}\label{UR3.3-new3.3}
{\bf{p}}.\psi(p)\Rightarrow p_{N}
\psi(p_{N})=\frac{\hbar}{(N_{p}-\frac{1}{4N_{p}})\ell}\psi(p_{N})\approx
\frac{\hbar}{N_{p}\ell}\psi(p_{N}),\nonumber
\\
{\bf{x}}.\psi(p)\Rightarrow{\bf{x}}.\psi(p_{N}) =\imath \hbar
(1+\frac{1}{(N_{p}-\frac{1}{4N_{p}})^{2}})
\Delta^{-1}_{p}\psi(p_{N})\approx  \nonumber
\\
\approx \imath \hbar(1+\frac{1}{N_{p}^{2}})
\Delta^{-1}_{p}\psi(p_{N}).
\end{eqnarray}

The scalar product $\langle \psi \vert \phi \rangle$ from \cite{Kempf}
\begin{equation} \label{sp}
\langle \psi \vert \phi \rangle  =  \int_{-\infty}^{+\infty}
\frac{dp}{1+ \beta p^2} \psi^{*}(p) \phi(p)
\end{equation}
in the case of {low energies} $1\ll |N_{\Delta p}|\leq
\widetilde{\textbf{N}}<\infty$ is replaced by the sum
\begin{eqnarray} \label{sp.1}
\langle \psi \vert \phi \rangle  =  \int_{-\infty}^{+\infty}
\frac{dp}{1+ \beta p^2} \psi^{*}(p) \phi(p)\Rightarrow \nonumber
\\\Rightarrow\langle \psi \vert \phi
\rangle_{1\ll|N_{p}|\leq
\widetilde{\textbf{N}}}=\sum_{1\ll|N_{p}|\leq
\widetilde{\textbf{N}}}\frac{\Delta_{p}(p_{N})\psi^{*}(p_{N})
\phi(p_{N})}{(1+\frac{1}{(N_{p}-\frac{1}{4N_{p}})^{2}})}\approx
\nonumber
\\ \approx\sum_{1\ll|N_{p}|\leq \widetilde{\textbf{N}}}
\frac{\Delta_{p}(p_{N})\psi^{*}(p_{N})
\phi(p_{N})}{(1+\frac{1}{N_{p}^{2}})}.
\end{eqnarray}

And since $|N_{p}|\gg 1$ is a variable, in fact $p_{N}$ is
continuously varying and, proceeding from the above formulae, we
can assume that to a high accuracy the function
$\phi(p_{N})$,($\psi^{*}(p_{N})$) is differentiable in terms of
this variable.

On the other hand, at {high energies}, when for $|N_{p}|\approx 1$
the presentation is fairly discrete, the scalar product Equation~\eqref{sp}
is replaced by the sum
\begin{eqnarray} \label{sp.1}
\langle \psi \vert \phi \rangle  =  \int_{-\infty}^{+\infty}
\frac{dp}{1+ \beta p^2} \psi^{*}(p) \phi(p)\Rightarrow \nonumber
\\\Rightarrow\langle \psi \vert \phi
\rangle_{|N_{p}|\approx 1}=\sum_{|N_{p}|\approx 1}
\frac{\Delta_{p}(p_{N})\psi^{*}(p_{N})
\phi(p_{N})}{(1+\frac{1}{(N_{p}-\frac{1}{4N_{p}})^{2}})}.
\end{eqnarray}

We consider only two cases: (a) $1\ll|N_{p}|\leq \widetilde{\textbf{N}}$,
 {``Quantum Consideration, Low Energies''}
and (b) $|N_{p}|\approx 1$,{``Quantum Consideration, High Energies''}. The case (c)
\begin{eqnarray} \label{sp.1.1}
\widetilde{\textbf{N}}\ll|N_{p}|<\infty
\end{eqnarray}
is omitted in this Section as it is associated with the {
``Classical Picture''}.

 Then at all the energy scales $\langle \psi \vert \phi \rangle_{N_{p}}$
may be formally represented as follows:
\begin{eqnarray} \label{sp.2}
\langle \psi \vert \phi \rangle_{N_{p}}=\langle \psi \vert \phi
\rangle_{1\ll|N_{p}|\leq \widetilde{\textbf{N}}}+\langle \psi
\vert \phi \rangle_{|N_{p}|\approx 1}.
\end{eqnarray}

However, with the formalism and terms proposed in this work, and
also with the use of the Formula~(\ref{GUP3}) that in this case
takes the form
\begin{equation}\label{GUP3.new1}
 (|N_{p}|\approx 1)\rightarrow (1\ll|N_{p}|\leq \widetilde{\textbf{N}}),
\end{equation}
it seems more logical to consider the two components in
Equation~\eqref{sp.2} separately, the first component  originating in the
process of the low-energy transition from the second component as~follows:
\begin{equation}\label{UR3.3-new4}
\langle \psi \vert \phi \rangle_{|N_{p}|\approx
1}\stackrel{|N_{p}|\gg 1}{\Rightarrow}\langle \psi \vert \phi
\rangle_{1\ll|N_{p}|\leq \widetilde{\textbf{N}}}.
\end{equation}

Clearly, the first part of formula (13) from \cite{Kempf} holds as
well in the general case for each of the components Equation~\eqref{sp.2}
\begin{equation}\label{sp.4}
\langle(\psi|{\bf{p}})|\phi\rangle =
\langle\psi|({\bf{p}}|\phi)\rangle
\end{equation}

The second part of formula (13) from \cite{Kempf} \begin{equation}\label{sp.5}
\langle(\psi|{\bf{x}})|\phi\rangle =
\langle\psi|({\bf{x}}|\phi)\rangle
\end{equation}
takes place (to a high accuracy)    for the {
low-energy} case $1\ll |N_p|\leq
\widetilde{\textbf{N}}<\infty$, \emph{i.e.}, for the first component in Equation~\eqref{sp.2}.

Indeed, in this case, due to the condition $|N_{p}|\gg 1$, we have
\begin{eqnarray}\label{UR3.3-new5}
\Delta_{p}p_{N}\approx dp;\Delta^{-1}_{p}\psi(p_{N})\approx
\partial_p\psi(p_{N})
\nonumber
\\\enskip or
\nonumber
\\
\
\lim\limits_{|N_{p}|\rightarrow\infty,(\widetilde{\textbf{N}}\rightarrow
\infty)}\Delta_{p}p_{N}=
dp;\lim\limits_{|N_{p}|\rightarrow\infty,(\widetilde{\textbf{N}}\rightarrow
\infty)}\Delta^{-1}_{p}\psi(p_{N})=
\partial_p\psi(p_{N}).
\end{eqnarray}

Then in this ({low-energy}) case there exists the analog of formula
(15) from \cite{Kempf}
\begin{eqnarray}\label{sp.6}
\langle\psi|({\bf{x}}|\phi)\rangle=\sum_{1\ll|N_{p}|\leq
\widetilde{\textbf{N}}-1}
\frac{\Delta_{p}(p_{N})}{(1+\frac{1}{N_{p}^{2}})}\psi^{*}(p_{N})i\hbar
(1+\frac{1}{N_{p}^{2}}){\Delta^{-1}_{p}}(\phi(p_{N}))=\nonumber
\\=\sum_{1\ll|N_{p}|\leq \widetilde{\textbf{N}}-1}
\Delta_{p}(p_{N})\psi^{*}(p_{N})i\hbar{\Delta^{-1}_{p}}(\phi(p_{N}))\approx\nonumber
\\\approx \langle(\psi|{\bf{x}})|\phi\rangle =
\sum_{1\ll|N_{p}|\leq \widetilde{\textbf{N}}-1}
\Delta_{p}(p_{N})(i\hbar{\Delta^{-1}_{p}}\psi(p_{N}))^{*}\phi(p_{N}).
\end{eqnarray}

It is important to note the following remarks:
\\(1) The operator ${\bf{x}}$  is defined in the case of low energies
only for the functional space $\phi(p_{N})_{1\ll|N_{p}|\leq
\widetilde{\textbf{N}}-1}$. Really, because of the existence of
the Formula (\ref{UR3.3-new2.1}), the extreme point $N_{p}$, (such
that $(N_{p}+1)(N_{p}+2)>\textbf{N}$) ``moves'' this operator
beyond the domain under study $\Delta p_{min}=\textmd{p}$.
Therefore, replacing $N_{p}\mapsto N_{p}+1,N_{p}+1\mapsto N_{p}+2$
in Formula (\ref{UR3.3-new2.2}), one can easily get the estimate
of $\widetilde{\textbf{N}}-1$ instead of  $\widetilde{\textbf{N}}$
as seen in Equation~\eqref{sp.6}.
\\
\\(2) Despite the fact that the operator ${\bf{x}}$ is also defined at high energies,
\emph{i.e.}, for  $\phi(p_{N})_{|N_{p}|\approx 1}$,in general the
property Equation~\eqref{sp.5} in this case has no place for lack
of Formulae (\ref{UR3.3-new5}).
\\
\\(3) In all the cases when we consider $|N_{p}|\gg
1$ (low energies) the ``cutoff'' for some upper bound
$p_{max}$,($p_{max}\ll P_{pl}$),$1\ll N_{p_{max}}<|N_{{p}}|,p\neq
p_{max}$ is determined by the initial conditions of the solved
problem.
\\
\\(4) It is clear that in the relativistic case $\Delta p_{min}=\textmd{p}$
leads to a {minimal variation in the energy}
\begin{equation}\label{Comm2}
|\Delta E_{min}|=(\Delta p)_{min}c=\frac{\textmd{p}}{\textbf{N}}c.
\end{equation}

(5) In this work a {minimal variation} of the momentum $\Delta
p_{min}$ has been introduced from the additional assumptions but,
as shown in \cite{Park}, a {minimal variation} of the momentum may
arise from the {Extended Uncertainty Principle (EUP)} as follows:
\begin{equation}\label{EUP}
 \Delta x_{i}\Delta p_{j}\geq \hbar \delta_{ij} [1 + \beta^{2}
\frac{(\Delta x_{i})^{2}}{\l^2}],
\end{equation}
where $l$ is the characteristic, large length scale $l\gg l_{p}$
and $\beta$ is a dimensionless real constant on the order of unity
\cite{Park}. From Equation~\eqref{EUP} we get an absolute minimum in the
momentum uncertainty
\begin{equation}\label{EUP0}
\Delta p_{i}\geq \frac{2\hbar \beta}{l}.
\end{equation}

~~~In \cite{Kim1} GUP and EUP are combined by the principle called
the {Symmetric Generalized Uncertainty Principle (SGUP)}:
\begin{equation}\label{SGUP1}
  \Delta x \Delta p \ge \hbar \left( 1 + \frac{(\Delta x)^2}{L^2} +
      \l^2 \frac{(\Delta p)^2}{\hbar^2} \right),
\end{equation}
where $l\ll L$ and $l$ defines the limit of the UV-cutoff (not
being such up to a constant factor as in the case of GUP).Then
\begin{center}
 $\Delta x_{\rm min} =
2l/\sqrt{1-4\l^2/L^2}=\ell$,
\end{center}
whereas $L$ defines the limit for IR-cutoff \emph{i.e.}, we have a
\begin{center}
$\Delta p_{\rm min} = 2\hbar/(L\sqrt{1-4\l^2/L^2)}$.
\end{center}

(6) Of course, this paper is only the first step to resolve the
Quantum Theory in terms of the {measurable quantities} using
{Definition 2}. It is necessary to study thoroughly the
{low-energy case} $E\ll E_P$ and the correct transition to high
energies $E\propto E_P$. The author is planning to treat these
problems in his further works.

\section {Gravity Markov's Model in Terms of Measurable Quantities}

This heuristic model was introduced in the work \cite{Mark1} at
the early eighties of the last century. This model already
considered by the author in his previous paper \cite{shalyt-IJMPD}
is treated from the standpoint of the above-mentioned arguments.
In \cite{Mark1}, it is assumed that ``by the universal decree of
nature a quantity of the material density $\varrho$ is always
bounded by its upper value given by the expression that is
composed of fundamental constants'' (\cite{Mark1}, p. 214):
\begin{equation}\label{Mark1}
\varrho\leq\varrho_{p}=\frac{c^{5}}{G^{2}\hbar},
\end{equation}
with $\varrho_{p}$ as ``Planck's density''.

Then the quantity
\begin{equation}\label{Mark4.1}
\wp_{\varrho}=\varrho/\varrho_{p}\leq 1
\end{equation}
 is the {deformation parameter} as it is used in \cite{Mark1} to
construct the following {of Einstein equations deformation or
$\wp_{\varrho}$-deformation} (Formula (2) in \cite{Mark1}):
\begin{equation}\label{Mark5}
R^{\nu}_{\mu}-\frac{1}{2}R\delta^{\nu}_{\mu}=\frac{8\pi
G}{c^{4}}T^{\nu}_{\mu}(1-\wp_{\varrho}^{2})^{n}-\Lambda\wp_{\varrho}^{2n}\delta^{\nu}_{\mu},
\end{equation}
where $n\geq 1/2$, $T^{\nu}_{\mu}$--energy-momentum tensor,
$\Lambda$---cosmological constant.

The case of the parameter $\wp_{\varrho}\ll 1$ or $\varrho\ll \varrho_{p}$
correlates with the classical Einstein equations, and the case
when $\wp_{\varrho}= 1$---with the de Sitter Universe. In this
way Equation~\eqref{Mark5} may be considered as $\wp_{\varrho}$-deformation
of the General Relativity.

As shown in \cite{shalyt-IJMPD},
$\wp_{\varrho}$-of Einstein equations deformation Equation~\eqref{Mark5} is
nothing else but $\alpha$-deformation of GR for the parameter
$\alpha_{l}$  at $a=l$ from Equation~\eqref{D1}.
\\If $\varrho=\varrho_{l}$ is the average material density for
the Universe  of the characteristic linear dimension $l$, \emph{i.e.}, of
the volume $V\propto l^{3}$, we have
\begin{equation}\label{Mark8}
\wp_{l,\varrho}=\frac{\varrho_{l}}{\varrho_{p}}\propto\alpha_{l}^{2}=\omega\alpha_{l}^{2},
\end{equation}
where $\omega$  is  some computable factor.

 Then it is clear that
$\alpha_{l}$-representation Equation~\eqref{Mark5} is of the form
\begin{equation}\label{Mark5.1}
R^{\nu}_{\mu}-\frac{1}{2}R\delta^{\nu}_{\mu}=\frac{8\pi
G}{c^{4}}T^{\nu}_{\mu}(1-\omega^{2}\alpha_{l}^{4})^{n}-\Lambda\omega^{2n}\alpha_{l}^{4n}\delta^{\nu}_{\mu},
\end{equation}
or in  the  general  form we have
\begin{equation}\label{Mark5.2}
R^{\nu}_{\mu}-\frac{1}{2}R\delta^{\nu}_{\mu}=\frac{8\pi
G}{c^{4}}T^{\nu}_{\mu}(\alpha_{l})-\Lambda(\alpha_{l})\delta^{\nu}_{\mu}.
\end{equation}

But, as r.h.s. of Equation~\eqref{Mark5.2} is dependent on $\alpha_{l}$ of
any value and particularly in the case $\alpha_{l}\ll 1$, \emph{i.e.},  at
$l\gg \ell$, l.h.s of Equation~\eqref{Mark5.2} is also dependent on
$\alpha_{l}$ of any value and Equation~\eqref{Mark5.2} may be written as
\begin{equation}\label{Mark5.2new}
R^{\nu}_{\mu}(\alpha_{l})-\frac{1}{2}R(\alpha_{l})\delta^{\nu}_{\mu}=\frac{8\pi
G}{c^{4}}T^{\nu}_{\mu}(\alpha_{l})-\Lambda(\alpha_{l})\delta^{\nu}_{\mu}.
\end{equation}

Thus, in this specific case we obtain the explicit dependence of
GR on the available energies $E\sim \frac{1}{l}$, that is
insignificant at low energies or for $l\gg \ell$ and, on the
contrary, significant at high energies, $l\rightarrow \ell$.

\subsection{Low  Energies, $E\ll E_{P}$}
\label{section6.1}

{\bf 1.} {Low energies. Nonmeasurable case}. In this case at low
energies, using Formula (\ref{D1}) in the limit $\ell=0$ for
$a=l$,  we get a {continuous theory} coincident with the General
Relativity.
\\
\\{\bf 2.} {Low energies. Measurable case}. In this case at
low energies, using Formulas (\ref{D1}) and (\ref{D1.new1}) for
\mbox{$\ell \neq 0$,} for $a=l$ (and hence for $N_{l}\gg 1$), we
get a {discrete theory} which is a {``nearly continuous theory''},
practically similar to the General Relativity with the slowly
(smoothly) varying parameter $\alpha_{l(t)}$,
 where $t$---time.
\\So, due to low energies and momentums $(E\ll E_{P},p\ll P_{Pl})$,
the {``continuous case''} {\bf 1} (General Relativity) and the
{``discrete case''} {\bf 2} that is actually a {``nearly
continuous case''}.

\subsection{High  Energies, $E\approx E_{P}$}
\label{section6.2} {At high energies we consider the measurable
case} only. Then it is clear that at high energies the parameter
$\alpha_{l(t)}$ is discrete and for the limiting value of
$\alpha_{l(t)}=1$  we get a discrete series of equations of the
form Equation~\eqref{Mark5.2new} (or a single equation of this
form met by a discrete series of solutions) corresponding to
$\alpha_{l(t)}=1;1/4;1/9;...$

As this takes place, $T^{\nu}_{\mu}(\alpha_{l})\approx 0$, and in
both cases as {\bf 2} in 6.1 as well as  \ref{section6.2}
$\Lambda(\alpha_{l})$ is not longer a cosmological  constant,
being a   dynamical cosmological term.

Note that because of Formula (\ref{GUP1.new5}) given in Section
\ref{Section3}, $\sqrt{\alpha_{l(t)}}$  in cases {\bf 2} in 6.1
and \ref{section6.2} is an element of the lattice $Lat_{P-E}$ from
Section \ref{Section4}. And in case {\bf 2} it is an element of
the sublattice $Lat_{P-E}[LE]$, whereas case \ref{section6.2} is
associated with the sublattice $Lat_{P-E}[HE]$.

It seems expedient to make some important remarks:
\\
\\{\bf (1)} In formulae (\ref{GUP3.new1}) and (\ref{UR3.3-new4}) of Section
\ref{Section5} in this work we have considered the transition
\begin{eqnarray}\label{Trans1}
{Quantum\enskip Theory\enskip in\enskip High\enskip
Energies\enskip (QTHE)} \Rightarrow \nonumber
\\ \Rightarrow{
 Quantum\enskip Theory\enskip in\enskip Low\enskip Energies\enskip (QTLE)}.
\end{eqnarray}

However, according to the modern knowledge, the (quantum) gravity
phase begins {only} at very high energies at Planck scales,
\emph{i.e.}, the case (a) from Section \ref{Section5} is inexistent, and hence it is
more correct to consider the transition
\begin{eqnarray}\label{Trans2}
{Quantum\enskip Theory\enskip in\enskip High\enskip
Energies\enskip (QTHE)} \Rightarrow \nonumber
\\ \Rightarrow{Classical\enskip Theory\enskip   (Low\enskip Energies)}.
\end{eqnarray}

And this corresponds to the case (c) that has been omitted from
consideration in Section \ref{Section5} Equation~\eqref{sp.1.1} with
$\widetilde{\textbf{N}}=1$
\begin{equation}\label{Trans3}
 (|N_{p}|\approx 1)\rightarrow (1\ll|N_{p}|< \infty).
\end{equation}

{\bf (2)} Generally speaking, as \ref{section6.2} and case {\bf 2}
in \ref{section6.1} are associated with {measurable cases} for
{low energies} and {high energies}, respectively, all the terms of
the Equation (\ref{Mark5.2new}): $
R^{\nu}_{\mu}(\alpha_{l}),R(\alpha_{l}),T^{\nu}_{\mu}(\alpha_{l}),\Lambda(\alpha_{l})$
must be expressed in terms of {measurable quantities} in view of
{Definition 2} from Section \ref{Section3}. But this problem still
remains to be solved. In fact, it is reduced to the construction
of the following {\it ``measurable''} {deformations} in the sense
of {Definition 2} in Section \ref{Section3} as follows:
\begin{eqnarray}\label{Trans4}
\lim\limits_{\ell\rightarrow 0}(R^{\nu}_{\mu}(\alpha_{l}\ll
1),R(\alpha_{l}\ll 1),T^{\nu}_{\mu}(\alpha_{l}\ll
1),\Lambda(\alpha_{l}\ll 1))\rightarrow\nonumber
\\
\rightarrow(R^{\nu}_{\mu},R,T^{\nu}_{\mu},\Lambda)
\end{eqnarray}
and
\begin{eqnarray}\label{Trans5}
\lim\limits_{(\alpha_{l}\approx 1)\rightarrow (\alpha_{l}\ll
1)}(R^{\nu}_{\mu}(\alpha_{l}\approx 1),R(\alpha_{l}\approx
1),T^{\nu}_{\mu}(\alpha_{l}\approx 1),\Lambda(\alpha_{l}\approx
1))\rightarrow \nonumber
\\ \rightarrow \lim\limits_{l_{min}\rightarrow
0}(R^{\nu}_{\mu}(\alpha_{l}\ll 1),R(\alpha_{l}\ll
1)\delta^{\nu}_{\mu},T^{\nu}_{\mu}(\alpha_{l}\ll
1),\Lambda(\alpha_{l}\ll 1)) \rightarrow\nonumber
\\ \rightarrow(R^{\nu}_{\mu},R,T^{\nu}_{\mu},\Lambda).
\end{eqnarray}

Here the first Equation~\eqref{Trans4} is a pure low-energy limiting
transition from the {measurable} variant of gravity to the
{nonmeasurable} one (or from a {discrete theory} to a {
continuous theory}), whereas the second Equation~\eqref{Trans5} from the
beginning is associated with the {measurable transition} from
high energies to low energies and then is coincident with the
first one.
\\
\\{\bf (3)} It  should  be  noted that  in \cite{Shalyt-AHEP2,Shalyt-JAP1}
in terms of {measurable quantities}, as an example, we have
studied gravity for the static spherically-symmetric horizon
space. It has been shown that, ``{...despite the absence of
infinitesimal spatial-temporal increments owing to the existence
of $l_{min}$ and the essential `discreteness' of a theory, this
discreteness at low energies is not `felt', the  theory in fact
being close to the original continuum theory. The indicated
discreteness is significant only in the case of high (Planck)
energies} '' \cite{Shalyt-AHEP2}. The Markov model considered in
this section represents the generalization of the above-mentioned
example. Of course, this model requires further thorough
investigation in terms  of {measurable quantities}.

\section{Conclusions}

\subsection{Measurable and  Non-Measurable Transitions in Gravity}
The illustration considered in the preceding Section
({Gravity
Markov's Model}) is {universal} considering the following:

First, using the formalism of this work, it is required to construct
a {measurable} deformation of the General Relativity (GR) at
{low energies} (Formula (\ref{Trans4})). This deformation is
denoted in terms \mbox{of $Grav[LE]^{\ell}$}
\begin{eqnarray}\label{Trans6}
Grav[LE]^{\ell}\stackrel{\ell \rightarrow 0}{\Rightarrow}GR.
\end{eqnarray}

Next, we should construct the high-energy deformation
(denoted in terms of $Grav[HE]^{\ell}$), this time for
$Grav[LE]^{\ell}$ (the first arrow in the Formula (\ref{Trans5}))
\begin{eqnarray}\label{Trans7}
Grav[HE]^{\ell}\stackrel{\alpha_{l}\rightarrow
0}{\Rightarrow}Grav[LE]^{\ell}.
\end{eqnarray}

At the present time the majority of the proposed approaches to
quantization of gravity are associated with the construction of
the following transition:
\begin{eqnarray}\label{Trans8}
GR{\Rightarrow}Grav[HE]^{\ell}.
\end{eqnarray}

But, by the author opinion, this is impossible. It seems that for
correct quantization of gravity one needs reversal of the arrow
from Equation~\eqref{Trans7}
\begin{eqnarray}\label{Trans9}
Grav[LE]^{\ell}(\alpha_{l}\approx 0,\alpha_{l}\neq
0)\stackrel{\alpha_{l}\rightarrow
1}{\Rightarrow}Grav[HE]^{\ell}(\alpha_{l}\approx 1).
\end{eqnarray}

The above results indicate that the low-energy {``measurable''}
gravity variant $Grav[LE]^{\ell}$ should be very close to GR but
different at the same time.

The author is hopeful that the correct construction of a
low-energy $Grav^{\ell}$ close to GR  allows for a more natural
transition to quantum (Planck) gravity. Besides, within the notion
of {measurability}, gravity could be saved from some odd
solutions, from wormholes in particular.

\subsection{Measurable and  Non-measurable Transitions in Quantum Theory}

The situation is similar for a quantum theory too. In the general case,
 based on the parameter $\alpha_{a}$ (Formula
(\ref{D1.new1}) of Section \ref{Section3}), this means that {there exists
the following correct limiting \mbox{high-energy transition}:}
\begin{equation}\label{Concl1.1}
\lim\limits_{\ell \neq 0,|N_{a}|\gg
1}\alpha_{a}\stackrel{High\enskip
Energy}{\Rightarrow}\lim\limits_{\ell \neq 0,|N_{a}|\approx
1}\alpha_{a}
\end{equation}
and {there is no correct limiting high-energy transition}
\begin{equation}\label{Concl2.1}
\lim\limits_{\ell=0}\alpha_{a}\stackrel{High\enskip
Energy}{\Rightarrow}\lim\limits_{\ell \neq 0,|N_{a}|\approx
1}\alpha_{a}.
\end{equation}

The first of them corresponds to the transition from a {
measurable} theory at low energies to a {measurable} theory at
high energies \begin{eqnarray}\label{Trans10}
QT[LE]^{\ell}\stackrel{N_{a}\rightarrow
1}{\Rightarrow}QT[HE]^{\ell}.
\end{eqnarray}

Whereas the second
\begin{eqnarray}\label{Trans10.1}
QT\stackrel{N_{a}\rightarrow 1}{\Rightarrow}QT[HE]^{\ell}
\end{eqnarray}
(here $QT[LE]^{\ell},QT[HE]^{\ell},QT$ are quantum theories with
the minimal length $\ell \neq 0$ at low energies $E\ll E_p$, at
high energies $E\approx E_p$, and the well-known (continuous)
quantum theory with $l_{min}=0$).

However, the whole theoretical physics, where presently at low energies $E\ll E_{P}$
the minimal length $\ell$ is not involved at all (\emph{i.e.},
$l_{min}=0$),  is framed around a search for the {nonexistent
limits} Equation~\eqref{Concl2.1} (correspondingly Equation~\eqref{Trans10.1}).

 Of course, in this case the low-energy {``measurable''}
variant $QT[LE]^{\ell}$ of $QT$ by its results will be very close
to the initial theory $QT$, as indicated in
\cite{Shalyt-AHEP2,Shalyt-JAP1},  and Section \ref{Section5} of the
present work. But these theories are different by nature: the
first of them is discrete and the second one is continuous.
Nevertheless, it is clear that the main requirement in this case
is associated with the {``Compatibility Principe''}:\vspace{6pt}

\noindent{\it at low energies the resolved variant $QT[LE]^{\ell}$ must,
to a high accuracy, represent the well-known approved  results of
the corresponding continuous theory $QT$}.\vspace{6pt}

These theories should be differing considerably at least
on going to high energies $E\approx E_p$.

The hypothesis set by the author is that correct construction of the {``measurable''}
transition to high energies (Formula (\ref{Trans10})) should
naturally lead to solution of the ultraviolet divergences problem
(initially in terms of the finite {measurable} quantities).

\subsection{Summary of 7.1 and 7.2 is {Such} \cite{Shalyt-JAP1}}\label{7.3}

{\bf 1.} When in the theory the minimal length $l_{min}\neq 0$ is
actualized (involved) at all the energy scales, a mathematical
apparatus of this theory must be changed considerably: no
infinitesimal space-time variations (increments) must be involved,
the key role being played by the definition of {measurability}
({Definition 2} from  Section \ref{Section3}).
\\
\\{\bf 2.} As this takes place the theory becomes {discrete} at all the energy
scales but at low energies (far from the Planck energies) the
sought for theory must be very close in its results to the
starting continuous theory (with $l_{min}=0$). In the process a
real {discreteness} is exhibited only at high energies which are
close to the Planck energies.
\\
\\{\bf 3.} By this approach the theory at low and high energies is
associated with a common single set of the parameters ($N_{L}$
from Formula (\ref{Introd 2.4}))  or with the dimensionless small
parameters ($1/N_{L}=\surd \alpha_{L}$) which are lacking if at
low energies the theory is continuous, \emph{i.e.}, when
$l_{min}=0$.

The principal objective of my further studies is to develop for
quantum theory and gravity, within the scope of the considerations
given in points {\bf 1}--{\bf 3}, the corresponding discrete
models~ (with~\mbox{$l_{min}~\neq~0$}) for all the energy scales
and to meet the following requirements:
\\
{\bf 4.} At low energies the models must, to a high accuracy,
represent the results of the corresponding continuous theories.
\\
{\bf 5.} The models should not have the problems of transition
from low to high energies and, specifically, the ultraviolet
divergences problem. By author's opinion, the problem associated
with points {\bf 4} and {\bf 5} is as follows.
\\
{\bf 6.} It is interesting to know why, with the existing
$l_{min}\neq 0,t_{min}\neq 0$ and discreteness of nature, at low
energies $E\ll E_{max}\propto E_{P}$ the apparatus of mathematical
analysis based on the use of infinitesimal space-time quantities
($dx_{\mu},\frac{\partial \varphi}{\partial x_{\mu}}$, and  so on)
is very efficient giving excellent results. The answer is simple:
in this case $l_{min}$ and $t_{min}$  are very far from the
available scale of $L$  and $t$, the corresponding $N_{L}\gg
1,N_{t}\gg 1$  being in general true but insufficient. There is a
need for rigorous calculations.

\begin{center}
{\bf Conflict of Interests}
\end{center}
The author declares that there is no conflict of interests
regarding the publication of this article.

\bibliographystyle{mdpi}

\begin{thebibliography}{----}

\bibitem{Shalyt-AHEP2}
Shalyt-Margolin, A.E. Minimal Length and the Existence of Some
Infinitesimal Quantities in Quantum Theory and Gravity. \emph{Adv.
High Energy Phys.} \textbf{2014}, \emph{2014}, 8, doi:10.1155/2014/195157.
%
%
\bibitem{Shalyt-JAP1}
Shalyt-Margolin, A.E. Holographic Principle, Minimal Length and
Measurability. \emph{J. Adv. Phys.} (In press)

\bibitem{Einst1}
Wald ,Robert. M. {\it General Relativity}; Chicago:The University
Chicago Press,USA,1984.

\bibitem{QG1}
Amelino-Camelia, G. Quantum Spacetime Phenomenology. {\it Living
Rev. Relativ.} \textbf{2013}, \emph{16}, 5--129.
%
%
\bibitem{Hawk-Penr1}
Penrose, R. {\it Quantum Theory  and  Space-Time}, Fourth Lecture in
{\it Stephen Hawking and Roger Penrose, The Nature  of  Space  and
Time}; Prinseton University Press: Princeton, NJ, USA, 1996.
%
%
\bibitem{Gar1}
 Garay, L. Quantum gravity and minimum length. {\it Int.
J. Mod. Phys. A} \textbf{1995}, \emph{10}, 145--146.
%
%
\bibitem{Planck2.1}
Amelino-Camelia, G.; Smolin, L. Prospects for constraining quantum
gravity dispersion with near term observations. {\it Phys. Rev. D}
\textbf{2009}, doi:10.1103/PhysRevD.80.084017.
%
%
\bibitem{Planck2.2}
 Gubitosi, G.; Pagano, L.; Amelino-Camelia, G.; Melchiorri, A.; Cooray, A. A constraint on
planck-scale modifications to electrodynamics with CMB
polarization data. {\it J. Cosmol. Astropart. Phys.} \textbf{2009}, \emph{908}, 21--34.
%
%
\bibitem{Planck2}
 Amelino-Camelia, G. Building a case for a planck-scale-deformed boost
action: The planck-scale particle-localization limit. {\it
Int. J. Mod. Phys. D} \textbf{2005}, \emph{14}, 2167--2180.
%
%
\bibitem{Planck3.1}
Hossenfelder, S.; Bleicher, M.; Hofmann, S.; Ruppert, J.; Scherer, S.; St\"{o}cker, H. Signatures in the Planck Regime. {\it Phys.
Lett. B} \textbf{2003}, \emph{575}, 85--99.
%
%
\bibitem{Planck3.2}
Hossenfelder, S. Running Coupling with Minimal Length. {\it
Phys. Rev. D} \textbf{2004}, doi:10.1103/PhysRevD.70.105003.
%
%
\bibitem{Planck3}
Hossenfelder, S. Self-consistency in Theories with a Minimal
Length. {\it Class. Quantum Gravity} \textbf{2006}, \emph{23}, 1815--1821.
%
%
\bibitem{Planck4}
Hossenfelder, S. Minimal Length Scale Scenarios for Quantum
Gravity. {\it Living Rev. Relativ.} \textbf{2013}, doi:10.12942/lrr-2013-2.
%
%
\bibitem{Heis1}
Heisenberg, W. Uber den anschaulichen Inhalt der
quantentheoretischen Kinematik und Mechanik. {\it Z.
Phys.} \textbf{1927}, \emph{43}, 172--198. (In German)
%
%
\bibitem{Mess}
Messiah, A. {\it Quantum Mechanics};  North Holland
Publishing Company: Amsterdam, The Netherlands, 1967; Volume 1.


\bibitem{Land2}
Berestetskii, V.B.; Lifshitz, E.M.; Pitaevskii, L.P. {\it
Relativistic Quantum Theory}; Pergamon: Oxford, UK, 1971.


\bibitem{Ven1}
 Veneziano, G.A. Stringy nature needs just two constants. {\it
Europhys. Lett.} \textbf{1986}, \emph{2}, 199--211.
%
%
\bibitem{Ven2}
Amati, D.; Ciafaloni, M.; Veneziano, G. Can spacetime be probed
below the string size? {\it Phys.~Lett.~B} \textbf{1989}, \emph{216},
41--47.

\bibitem{Ven3}
Witten, E. Reflections on the fate of spacetime. {\it Phys.
Today} \textbf{1996}, \emph{49},  24--28.
%
%
\bibitem{Polch}
Polchinski, J. {\it String Theory}; Cambridge
University Press: Cambridge, UK, 1998.


\bibitem{GUPg1}
Adler, R.J.; Santiago, D.I. On gravity and the uncertainty
principle. {\it Mod. Phys. Lett. A}  \textbf{1999}, \emph{14},
1371--1378.
%
%
\bibitem{Ahl1}
Ahluwalia, D.V. Wave-particle duality at the Planck scale:
Freezing of neutrino oscillations. {\it Phys. Lett. A} \textbf{2000},
\emph{275}, 31--35.

\bibitem{Ahl}
Ahluwalia, D.V. Interface of gravitational and quantum realms.
{\it Mod. Phys. Lett. A} \textbf{2002}, \emph{17},
1135--1145.
%
%
\bibitem{Magg1}
Maggiore, M. The algebraic structure of the generalized uncertainty
principle. {\it Phys. Lett. B} \textbf{1993}, \emph{319}, 83--86.

%
%
\bibitem{Magg2}
Maggiore, M. Black Hole Complementarity and the Physical Origin of
the Stretched Horizon. {\it Phys. Rev. D} \textbf{1994}, \emph{49},
2918--2921.



\bibitem{Magg3}
Maggiore, M. A Generalized Uncertainty Principle in Quantum
Gravity. {\it Phys. Rev. D} \textbf{1993}, doi:10.1016/0370-2693(93)91401-8.
%
%
\bibitem{Capoz}
Capozziello, S.; Lambiase, G.; Scarpetta, G. The Generalized
Uncertainty Principle from Quantum Geometry. {\it
Int. J. Theor. Phys.} \textbf{2000}, \emph{39}, 15--22.
%
%
\bibitem{Kempf}
Kempf, A.; Mangano, G.; Mann, R.B. Hilbert space representation
of the minimal length uncertainty relation. {\it Phys. Rev.
D} \textbf{1995}, \emph{52}, 1108--1118.
%
%
\bibitem{Nozari}
Nozari, K.; Etemadi, A.  Minimal length, maximal momentum and
Hilbert space representation of quantum mechanics.  {\it Phys. Rev.
D} \textbf{2012}, doi:10.1103/PhysRevD.85.104029.
%
%
\bibitem{shalyt1}
Shalyt-Margolin, A.E.; Suarez, J.G. Quantum Mechanics of the Early
Universe and Its Limiting Transition. Available online:
http://arxiv.org/abs/gr-qc/0302119 (accessed on 30 August 2003).


\bibitem{shalyt2}
Shalyt-Margolin, A.E.; Suarez, J.G. Quantum mechanics at
Planck scale and density matrix. {\it Int. J.
Mod. Phys. D} {\bf2003},  \emph{12}, 1265--1278.
%
%
\bibitem{shalyt3}
Shalyt-Margolin, A.E.; Tregubovich, A.Y.  Deformed density matrix
and generalized uncertainty relation in thermodynamics. {\it Mod.
Phys. Lett. A}  \textbf{2004}, \emph{19}, 71--82.
%
%
\bibitem{shalyt4}
Shalyt-Margolin, A.E. Non-unitary and unitary transitions in
generalized quantum mechanics, new small parameter and information
problemsolving. {\it Mod. Phys. Lett. A} \textbf{2004}, \emph{19},
391--403.
%
%
\bibitem{shalyt5}
Shalyt-Margolin, A.E. Pure states, mixed states and Hawking
problem in generalized quantum mechanics. {\it Mod. Phys.
Lett. A}   \textbf{2004}, \emph{19}, 2037--2045.
%
%
\bibitem{shalyt6}
Shalyt-Margolin, A.E. The universe as a nonuniform lattice in
finite-volume hypercube: I. Fundamental definitions and particular
features. {\it Int. J. Mod. Phys. D}
\textbf{2004}, \emph{13}, 853--864.
%
%
\bibitem{shalyt7}
Shalyt-Margolin, A.E. The Universe as a nonuniformlattice in the
finite-dimensional hypercube. II. Simple cases of symmetry
breakdown and restoration. {\it Int. J. Mod. Phys. A} \textbf{2005}, \emph{20}, 4951--4964.
%
%
\bibitem{shalyt9}
Shalyt-Margolin, A.E. The density matrix deformation in physics of
the early universe and some of its implications. In {\it Quantum
Cosmology Research Trends};  Reimer, A., Ed.; Nova~Science:
Hauppauge, NY, USA, 2005; pp. 49--92.
%
\bibitem{Fadd}
Faddeev, L. Mathematical view of the evolution of physics. {\it
Priroda} \textbf{1989}, \emph{5}, 11--16.
%
%
\bibitem{Land1}
Landau, L.D.; Lifshits, E.M. {\it Field Theory}; Theoretical
Physics:  Moskow, Russia, 1988; Volume 2.
%
%
\bibitem{Hoss}
Hossenfelder, S. Minimal Length Scale Scenarios for Quantum
Gravity. \emph{Living Rev. Relativ.} {\textbf{2013}}, {\em 36}, 2, doi:10.12942/lrr-2013-2.
%
%
\bibitem{Tawf}
Tawfik, A.N.; Diab, A.M. Generalized Uncertainty Principle:
Approaches and Applications. \emph{Int. J. Mod. Phys.~D} {\bf
2014}, {\em 23}, 1430025.
%
%
\bibitem{Vagen}
Vagenas, E.C.; Majhi, B.R. Modified Dispersion Relation, Photon's
Velocity, and Unruh Effect. \emph{Phys. Lett. B.} {\textbf{2013}}, {\em
725}, 477--483.
%
%
\bibitem{Nozar1}
Nozari, K.; Sefiedgar, A.S. Comparison of Approaches to Quantum
Correction of Black Hole Thermodynamics. \emph{Phys. Lett. B} {\bf
2006}, {\em 635}, 156--160.
%
%
\bibitem{Nozar2}
Nozari, K.; Fazlpour, B. Generalized Uncertainty Principle, Modified
Dispersion Relations and Early Universe Thermodynamics.
\emph{Gen. Relativ. Gravit.} {\textbf{2006}}, {\em 38}, 1661--1679.
%
%
\bibitem{shalyt-entropy2}
Shalyt-Margolin, A.  Entropy in the present and early universe: New
small parameters and dark energy problem. {\it Entropy} \textbf{2010}, \emph{12},
 932--952.
%
%
\bibitem{shalyt-IJMPD}
Shalyt-Margolin, A.E.  Quantum theory at planck scale, limiting
values, deformed gravity and dark energy problem. {\it
Int. J. Mod. Phys. D} \textbf{2012}, \emph{21},
doi:10.1142/S0218271812500137.
%
%
\bibitem{Park}
 Park, M.I. The Generalized Uncertainty Principle in (A)dS Space and the
 Modification of Hawking Temperature from the Minimal Length. \emph{Phys. Lett. B} {\textbf{2008}},
 {\em 659}, 698--702.
%
%
\bibitem{Kim1}
Kim, W.; Son, E.J.; Yoon, M.
Thermodynamics of a black hole based on a generalized uncertainty
principle. \emph{JHEP} {\textbf{2008}}, {\em 8}, doi:10.1088/1126-6708/2008/01/035.
%
%
\bibitem{Mark1}
 Markov, M.A. Ultimate Matter  Density  as the Universal Low
 of Nature. {\it JETP Lett.} \textbf{1982}, \emph{36}, 214--216.
%
%
%

\end{thebibliography}
\renewcommand\bibname{References}

\end{document}